\RequirePackage{fix-cm}
\pdfoutput=1
\documentclass[smallextended]{svjour3} 

\usepackage[T1]{fontenc}
\usepackage[utf8]{inputenc}
\usepackage{graphicx}
\usepackage{cite}
\usepackage[normalem]{ulem}
\usepackage{amsmath,amssymb,amsfonts,mathtools,bm,mathrsfs}
\usepackage{xcolor}

\newcommand{\be}{\begin{equation}}
\newcommand{\ee}{\end{equation}}
\newcommand{\bea}{\begin{eqnarray}}
\newcommand{\eea}{\end{eqnarray}}

\newcommand{\erf}{\mathrm{erf}}

\newcommand{\disavg}[1]{\left\langle #1\right\rangle}

\begin{document}

\title{Localization induced by Sinai disorder in mass-exchange models}
\author{Michele Giusfredi$^{1,2,3}$, Luca Avena$^{4}$ and Paolo Politi$^{2,3}$}
\institute{Michele Giusfredi \at Michele.Giusfredi@unifi.it
           \and Luca Avena \at Luca.Avena@unifi.it
	\and Paolo Politi \at Paolo.Politi@cnr.it 
\and
              $^1$ Dipartimento di Fisica e Astronomia, Universit\`a di Firenze,
via G. Sansone 1 I-50019, Sesto Fiorentino, Italy \at
\and
$^2$ Istituto dei Sistemi Complessi, Consiglio Nazionale
delle Ricerche, via Madonna del Piano 10, I-50019 Sesto Fiorentino, Italy \at
\and
$^3$ Istituto Nazionale di Fisica Nucleare, Sezione di Firenze,
via G. Sansone 1 I-50019, Sesto Fiorentino, Italy \at
\and
$^4$ Dipartimento di Matematica e Informatica `Ulisse Dini',
Universit\`a di Firenze, Viale Morgagni 67/a I-50134, Firenze, Italy
}
\date{Received: \today}
\authorrunning{M. Giusfredi, L. Avena, P. Politi}
\titlerunning{Localization induced by Sinai disorder}
\maketitle

\begin{abstract}
We introduce a conservative continuous-mass-exchange model 
with quenched disorder in the splitting laws on the bonds of a finite segment. 
Although the dynamics is generically irreversible, its stationary mean mass profile remarkably turns out to be exactly solvable through an associated birth-and-death chain and has a Gibbs representation in terms of a simple random-walk potential. 
If $N$ is the number of bonds and $\beta$ measures the strength of disorder,
the parameter $\beta\sqrt N$ separates an asymptotically flat profile, a Brownian crossover, 
and a profile localized near the global minimum. 
We characterize the fluctuations of macroscopic masses and identify arcsine laws for the localization centre, the competition between left and right valleys, and the disorder-averaged mass measure.
Furthermore, for the extremal masses, we discuss the emergent limiting random variable in terms of the literature linking finite RW paths seen from its minimum to infinite-horizon walks conditioned to stay positive.
Finally, in proximity of the localization transition point $\beta=0$, we show that  
the inverse participation ratio for the extremal masses give rises to a scaling with decay of order  $k^{-3/2}$ consistent with the scaling limit of the latter conditioned walk.
	In the same limit, the inverse participation ratio shows a scaling form.

\end{abstract}

\section{Introduction}
\label{sec:introduction}

Stochastic mass-exchange models provide a natural class of generalized interacting particle systems for studying the redistribution and transport of a conserved quantity. Despite their simple local dynamic rules, these models present a delicate mathematical structure and even their stationary states may display a rich structure. Recent advances in duality, intertwining, solvable stationary states and spectral methods have substantially broadened our understanding of these systems under proper reversible conditions~\cite{giardina2009,carinci2013,giardina2025,kim2025,caputo2025,capanna2024,carinci2024}.
In this work we introduce a quenched, bond-disordered generalization of the Kipnis--Marchioro--Presutti (KMP) model~\cite{kmp1982} on a finite one-dimensional segment. The total mass on each selected bond is redistributed according to a Beta splitting law, as in the general reversible exchange models mentioned above, though the associated parameters depend on the extra disorder attached to the bonds and remain fixed throughout the evolution.
A distinctive feature is that the resulting stationary mean mass profile remains exactly solvable, although the mass-exchange dynamics is generically irreversible. Its first-moment equations close and identify this profile as the invariant probability measure of an inhomogeneous birth-and-death chain in a random environment. The resulting product formula defines an effective random potential, which becomes a symmetric random walk for the balanced disorder considered here. Thus a Sinai-type landscape emerges naturally from the exchange mechanism, connecting conservative mass redistribution with localization, random-walk extrema and conditioned paths. This explicit representation allows us to use results from fluctuation theory and random walk invariance principles to describe both localization and its macroscopic fluctuations~\cite{bolthausen1976,bertoin1993,bertoin1994,caravenna2005,caravenna2008}.
This novel framework is linked to a one-dimensional recurrent random walks in random environment.  The relevant potential barriers are in fact those underlying Sinai--Golosov localization and the limiting infinite-valley description~\cite{sinai1982,golosov1984,kesten1986}.
As mentioned, the central simplification is that the quenched stationary mean profile can be computed exactly.  If $a_i(\omega)$ denotes the stationary fraction of mass at site $i$ in the environment $\omega$, for closed boundary conditions, and $N+1$ is the system size, then 
\begin{equation}
\label{eq:central_gibbs_profile}
a_i(\omega)=\frac{e^{-\beta V_i(\omega)}}{Z_N(\beta,\omega)},
\qquad
Z_N(\beta,\omega)=\sum_{j=0}^{N}e^{-\beta V_j(\omega)}.
\end{equation}
Here $V=(V_i)_{i=0}^N$ is a symmetric random walk in the spatial coordinate and $\beta\in [0,\infty)$ measures the strength of 
the bond disorder and governed the local asymmetry in the mass splitting mechanism, see Section \ref{sec:model}. This exact mapping in finite volume allows us to study localization using fluctuation theory for random walks and Brownian motion.

The competition between disorder and system size is controlled by $\beta\sqrt N$.  When $\beta\sqrt N\ll1$, the potential changes little on the thermal scale and the profile is nearly homogeneous.  When $\beta\sqrt N=O(1)$, the discrete Gibbs profile converges to a random density constructed from Brownian motion.
Whereas for  $\beta\sqrt N\gg1$ and in particular in the growing $N$ limit at fixed $\beta>0$, potential fluctuations dominate entropy and the mass becomes localized in a microscopic neighbourhood of the global minimum set.  We select the first minimizer and write
\begin{equation}
\label{eq:istar_definition_intro}
i_\star=\operatorname*{arg\,min}_{0\leq i\leq N}V_i
\end{equation}
where the first occurrence is chosen when the minimum is attained more than once.

Our second main result beyond the polymer like representation in Eq.~\eqref{eq:central_gibbs_profile} concerns macroscopic fluctuations.  
For a fixed observation point $x=i/N\in[0,1]$ and fixed $\beta>0$,
\begin{equation}
\label{eq:intro_W_limit}
-\frac{\log a_{\lfloor xN\rfloor}}{\beta\sqrt N}
\xrightarrow[N\to\infty]{d}
W(x)=\max\{\sqrt{x}H_1,\sqrt{1-x}H_2\},
\end{equation}
where $H_1,H_2$ are independent standard half-Gaussian variables.  The two terms measure the depth of the best valley respectively to the left and to the right of the observation point.  Thus Eq.~\eqref{eq:intro_W_limit} gives a direct probabilistic description of the exponentially small mass observed macroscopically far from the localization centre.

Three arcsine law then appears at different but connected levels.  First, the rescaled localization centre $i_\star/N$ converges to the (a.s. unique) time when the Brownian motion on $[0,1]$ reaches its minimum, corresponding to the classical arcsine L\'evy law \cite{levy1940}.  Second, the probability that the left contribution determines $W(x)$ is
\begin{equation}
\mathbb P\bigl(W(x)=\sqrt{x}H_1\bigr)
=\frac{2}{\pi}\arcsin\sqrt{x},
\end{equation}
which is exactly the arcsine distribution function.  Third, after averaging over the disorder, the normalised macroscopic mass measure converges to the arcsine density.  At the local level, see Fig.~\ref{fig:arcsin_profile}, this leads to
\begin{equation}\label{AvArcsin}
N\disavg{a_{\lfloor xN\rfloor}}
\longrightarrow \frac{1}{\pi\sqrt{x(1-x)}},
\qquad x\in(0,1),
\end{equation}
of which we will prove a slightly weaker version, see Eq.~\eqref{eq:arcsine_limit_profile} in Section \ref{sec:trapping}.

The three arcsine identities are tied to the geometry of the finite segment with closed boundaries.  The two ends distinguish the left and right portions of the potential and make the macroscopic position of its minimum non-uniform.  Analogous critical-scaling and localization statements continue to hold for a periodic geometry, after replacing the potential by the appropriate cyclic landscape, see Appendix~\ref{app:periodic}, but translation invariance removes the boundary-selected arcsine profile.  This should be contrasted with dynamical arcsine laws in strongly inhomogeneous diffusion and trap models.  In Bouchaud-type trap models the arcsine distribution describes aging through stable subordinators and clock processes~\cite{fontes2002,benarous2006}; here it arises instead as a static spatial law for the localization centre and the disorder-averaged stationary mass.

The paper is organized as follows.  Section~\ref{sec:model} introduces the exchange dynamics and derives the Gibbsian representation for the marginal masses at equilibrium at a given site.  Section~\ref{sec:partition_function} explores the structure of the resulting partition function and identifies the emerging phase transition and the associated regimes.  Section~\ref{sec:marginal_mass} establishes the macroscopic fluctuation law.  Section~\ref{sec:localization} studies the extremal mass at a deepest site in the different regimes.  Section~\ref{sec:trapping} is devoted to the three appearances of the arcsine law.  
Section~\ref{sec:numerics} offers numerical tests to check the arcsine profile,
to provide further insights into the mass profiles and into the inverse participation ratio.
Technical arguments are presented in the appendices.

\section{The model: stochastic exchange masses with bond Sinai disorder}
\label{sec:model}

Consider the segment with vertex set $\{0,\ldots,N\}$ , edge set $E=\{(i,i+1):0\le i<N\}$ and let $c=(c_0,\ldots,c_N)$ be a vector/configuration of non-negative masses with constant total mass associated to the vertex set of the graph, i.e.  a configuration $c$ belongs to the simplex
\[
\Omega_A=\{c\in[0,\infty)^{N+1}:\textstyle\sum_{i=0}^N c_i=A\},
\]
for some positive $A$>0. Such a number represents the conserved total mass of the system which without loss of generality we assume to be $A=1$.
To each edge we associate a quenched pair of positive parameters $(\alpha_{1,i},\alpha_{2,i})$ 
from a given collection  $\vec{\alpha}:=\{ (\alpha_{1,i},\alpha_{2,i}) \in (0,\infty)^2 : 0 \leq i<N \}$.
 Whenever edge $(i,i+1)$ is updated, a fresh splitting variable $r$ is sampled from a Beta distribution $\mathrm{Beta}(\alpha_{1,i},\alpha_{2,i})$, that is the random variable on $(0,1)$  with density
\[
g_i(r):=\frac{r^{\alpha_{1,i}-1}(1-r)^{\alpha_{2,i}-1}}{\mathrm B(\alpha_{1,i},\alpha_{2,i})},\qquad 0<r<1.
\]
The update configuration $T_{i,r}c$ leaves all other coordinates unchanged and sets
\begin{equation}
c'_i=r(c_i+c_{i+1}),\qquad c'_{i+1}=(1-r)(c_i+c_{i+1}).
\end{equation}
Each edge has an independent Poisson clock of rate $\lambda_i>0$ which to ease the exposition we set equal to $1$ unless stated otherwise. Formally, the generator describing the infinitesimal transitions of this continuous-time Markov chain on $
\Omega_A$ acts on bounded measurable functions $f:\Omega_A\to\mathbb R$ as
\begin{equation}\label{Generator}
(\mathcal L_{\vec{\alpha}}f)(c)=\sum_{i=0}^{N-1}\lambda_i\int_0^1\bigl[f(T_{i,r}c)-f(c)\bigr]g_i(r)\,dr.
\end{equation}
The dynamics above is the Beta redistribution generator used for generalized KMP models such as in the recent~\cite{giardina2009,carinci2013,giardina2025,kim2025,caputo2025,capanna2024,carinci2024} but with parameters attached here to edges rather than vertices.

It is worth stressing that the latter distinction matters for reversibility. In fact, if there are positive parameters $\theta_0,\ldots,\theta_N$ on the vertices such that $(\alpha_{1,i},\alpha_{2,i})=(\theta_i,\theta_{i+1})$ on every edge, each edge update samples its conditional Beta law and the Dirichlet law with parameters $(\theta_i)$ is known to be reversible, see e.g.~\cite{kim2025}. Considering arbitrary edge parameters as in the dynamics encoded in \eqref{Generator} need not satisfy this compatibility condition, and the resulting dynamics is irreversible. 

Nevertheless, the stationary mean profile turns out to be explicitly computable in terms of the chosen collection $\vec{\alpha}=\{ (\alpha_{1,i},\alpha_{2,i}) : 0 \leq i<N \}$ for the Beta-splitting parameters. Indeed, as shown in Appendix~\ref{app:partition_and_marginal}, writing $a_i=\mathbb E_{\vec{\alpha}}[c_i]$ for the steady mass at site $i$ with fixed chosen splitting parameters $\vec{\alpha}$ gives
\begin{equation}
a_0=\frac1Z,\qquad a_i=\frac{\prod_{k=0}^{i-1}\gamma_k}{Z},\qquad 1\le i\le N,
\label{eq:stationary_mean_profile}
\end{equation}
where
\begin{equation}
\gamma_i=\frac{1-\overline{r_i}}{\overline{r_i}}=\frac{\alpha_{2,i}}{\alpha_{1,i}},\qquad
\overline{r_i}=\frac{\alpha_{1,i}}{\alpha_{1,i}+\alpha_{2,i}},
\label{eq:gamma_definition}
\end{equation}
and
\begin{equation}
Z=1+\sum_{i=1}^N\prod_{k=0}^{i-1}\gamma_k.
\label{eq:stationary_normalization}
\end{equation}
Note that only the splitting averages $\overline{r_i}$'s enter this formula. In particular, such steady marginal masses can be seen associated to a reversible measure of a birth-and-death 
continuos time Markov chain on the segment with rates $q(i,i+1)=\lambda_i(1-\overline{r_i})$ and $q(i+1,i)=\lambda_i\overline{r_i}$. Its invariant probability vector is then precisely $(a_i)_{i<N}$, since $a_iq(i,i+1)=a_{i+1}q(i+1,i)$.

For the disorder studied below, fix two positive parameters $0<\alpha_1\le\alpha_2$ and let $\omega=(\omega_0,\ldots,\omega_{N-1})$ be a vector of $N$ independent \emph{spin} variables $w_i\in\{+1,-1\}$ attached to the bonds $(i,i+1)$. Set
\[
(\alpha_{1,i},\alpha_{2,i})=
\begin{cases}(\alpha_1,\alpha_2),& \text{if } \omega_i=+1,\\
(\alpha_2,\alpha_1),&  \text{if } \omega_i=-1.
\end{cases}
\]
Thus, if $G$ is the $\mathrm{Beta}(\alpha_1,\alpha_2)$ density, $g_i(r)=G(r)$ or $G(1-r)$ according to the sign $w_i$. In particular the first equation in \eqref{eq:gamma_definition} reads as $\gamma_i=\gamma^{\omega_i}$ with the one parameter $\gamma:=\alpha_2/\alpha_1\ge1$. Equivalently, fixing an environment $w$, in this one parameter case, the generator in Eq.\eqref{Generator} reads as \begin{equation*}
(\mathcal L_\omega f)(c)=\sum_{i=0}^{N-1}\lambda_i\int_0^1
[f(T_{i,r}c)-f(c)]\left[\frac{1+\omega_i}{2}G(r)+\frac{1-\omega_i}{2}G(1-r)\right]dr.
\end{equation*}
Defining $V_0=0$, $V_i=-\sum_{k=0}^{i-1}\omega_k$ and $\beta=\log\gamma$, we obtain
\begin{equation}
a_i(\omega)=\frac{\gamma^{-V_i}}Z=\frac{e^{-\beta V_i}}Z,
\label{eq.ai}
\end{equation}
with
\begin{equation}
Z=\sum_{i=0}^N\gamma^{-V_i},\qquad \omega\in\{\pm1\}^N.
\end{equation}
We will in particular focus on the balanced case where the spins variables $w_i$'s are independent and have value $\pm 1$ symmetrically with probability $1/2$. Thus the resulting potential $V_i$ above is a simple symmetric random walk in the spatial coordinate. Equation~\eqref{eq.ai} is a Gibbs representation of the stationary \emph{mean} masses but it does not identify the full stationary law of the mass-exchange process. Furthermore, we stress that the associated birth-and-death chain is reversible even when the mass-exchange process is not.

For $\beta=0$ ($\gamma=1$), $a_i=1/(N+1)$; the original uniform KMP splitting is recovered when $\alpha_1=\alpha_2=1$. As $\beta\to\infty$ at fixed $N$, the mean mass is shared equally among the sites attaining the global minimum of $V$. The next section examines how homogeneity is lost when disorder strength $\beta$ and system size $N$ vary.

\section{Partition function and disorder scaling}
\label{sec:partition_function}

The normalization factor in the Gibbs measure \eqref{eq:central_gibbs_profile} defines the partition function of the system:
\begin{equation}
\label{eq:partition_function_def}
Z_N(\beta) = \sum_{j=0}^N e^{-\beta V_j}
\end{equation}
The competition between disorder and system size is governed by the single parameter $\beta\sqrt N$.  The following emerging three regimes are naturally distinguished.
\begin{itemize} \item \emph{Localization}:
 in the supercritical regime,
\[
\beta\sqrt N\longrightarrow\infty,
\]
the fluctuations of the random potential dominate the normalization and the Gibbs mass is controlled by the extreme valleys; this includes, in particular, every fixed $\beta>0$ and it is what we refer to as the localization regime.
 \item \emph{Brownian macroscopic limit}:
on the critical line,
\[
\beta\sqrt N\longrightarrow\kappa\in(0,\infty),
\] the rescaled potential converges to a Brownian landscape and the stationary profile has a non-trivial macroscopic random field.  
\item  \emph{Flat homogeneous masses}:
finally, in the subcritical regime,
\[
\beta\sqrt N\longrightarrow0,
\]
the disorder becomes negligible on the macroscopic scale and the stationary profile approaches the flat profile of the homogeneous KMP model. 
\end{itemize}

Let us start describing the emergent limits of the partition function in the critical and supercritical regimes; the subcritical case recovers the flat mean profile of the homogeneous model, which follows directly from the fact that $\max_{0\le j\le N}|\beta_N V_j|\to0$ in probability.

Let $\beta_N=\kappa/\sqrt{N}(1+o(1))$ with $\kappa>0$. As $N \to \infty$, the rescaled partition function $N^{-1} Z_N(\beta)$ converges weakly to a functional of a standard Brownian motion $B_t$ on $[0,1]$:
\begin{equation}
\label{eq:partition_critical_limit}
\frac{1}{N} Z_N\left(\frac{\kappa}{\sqrt{N}}\right) \xrightarrow[N \to \infty]{d} \int_0^1 e^{-\kappa B_t} dt,
\end{equation}
which shows the emergence of the above mentioned Brownian macroscopic limit, see Appendix \ref{app:CriticalPartition}.

When we instead consider $\beta > 0$ to be fixed and send $N$ to infinity,
due to the recurrence properties of Sinai's potential, $Z_N(\beta)$ is overwhelmingly dominated by the extreme fluctuations of the potential valleys.
In particular, the partition function is dominated by the neighbourhood of the global minimum $m_N:=\min_{0 \le j \le N} V_j=V_{i_\star}$, recall Eq. \eqref{eq:istar_definition_intro}. Thus, the path-decomposition of the random walk viewed from this minimum yields
\begin{equation}
\label{eq:partition_fixed_beta_limit}
Z_N(\beta) e^{\beta m_N} 
=1 +\sum_{j=0}^{i_\star-1} e^{-\beta(V_j-V_{i_\star})} + \sum_{j={i_\star}+1}^N e^{-\beta(V_j-V_{{i_\star}})}
\xrightarrow[N \to \infty]{d} \mathcal{Z}_\infty^{(1)}(\beta) + \mathcal{Z}_\infty^{(2)}(\beta)
\end{equation}
where $\mathcal{Z}_\infty^{(1)}(\beta)$ and $\mathcal{Z}_\infty^{(2)}(\beta)$ are the exponential series of the two limiting walks seen from the selected minimum. See Appendix \ref{app:limitExtremeMass}.

Furthermore, the extreme values of the random walk potential $V_j$ dominate the behavior of $Z_N(\beta)$ in the supercritical regime.
Let $m_N=\min_{0\le j\le N}V_j$ denote the global minimum of the potential.

The role of the deepest valley is already visible from the elementary $\omega$-almost sure bounds
\begin{equation}
\label{eq:partition_bounds}
e^{-\beta m_N} \le Z_N(\beta) \le (N+1)e^{-\beta m_N}
\end{equation}
which imply, due to the classical diffusive scaling limit of $m_N$,
that for fixed $\beta>0$
$$\frac{1}{\beta\sqrt{N}} \log Z_N(\beta)  \xrightarrow[N\to\infty]{d}
H, $$
where the above denotes convergence in distribution, and $H$ stands for the half-Gaussian standard variable with density as in Eq. \eqref{eq:half_gaussian_density} characterizing the (a.s. unique) minimum of a Brownian motion in $[0,1]$.
Even for finite $N$, in the low-temperature limit $\beta \to \infty$, the bounds in \eqref{eq:partition_bounds} immediately gives
the convergence of the free energy density :
\begin{equation}
\lim_{\beta \to \infty} \frac{1}{\beta} \log Z_N(\beta) = -m_N = \max_{0\le j\le N}\sum_{k=0}^{j-1}\omega_k.
\end{equation}

\paragraph{Energy-level representation in finite volume.}
When the system size $N$ is finite, an alternative insightful route is to group the sites according to their height above the global minimum. That is, for every fixed $N$, define the number of site at level $r$ from the minimum as
\begin{equation}
\label{eq:energy_level_counts}
G_{N,r}:=\#\{0\le j\le N:V_j-m_N=r\},\qquad r\ge0.
\end{equation}
Then, $\omega$-almost surely and for every $\beta>0$,
\begin{equation}
\label{eq:partition_energy_levels}
e^{\beta m_N}Z_N(\beta)=\sum_{r\ge0}G_{N,r}e^{-\beta r}.
\end{equation}
This identity immediately gives the two following finite-size expansions at high and low temperature. As $\beta\downarrow0$ (high-temperature),
\begin{equation}
\label{eq:partition_energy_high_temperature_expansion}
e^{\beta m_N}Z_N(\beta)
=(N+1)-\beta A_{1,N}+\frac{\beta^2}{2}A_{2,N}+O(\beta^3),
\qquad A_{p,N}:=\sum_{r\ge0}r^pG_{N,r}.
\end{equation}
Thus the weak-disorder expansion probes successive moments of the energy-level profile. On the other hand, as $\beta\to\infty$ (low-temperature),
\begin{equation}
\label{eq:partition_energy_low_temperature_expansion}
e^{\beta m_N}Z_N(\beta)
=G_{N,0}+G_{N,1}e^{-\beta}+G_{N,2}e^{-2\beta}+O(e^{-3\beta}),
\end{equation}
so this high-disorder expansion resolves the valley level by level, starting from the degeneracy of the global minimum. Consequently, knowledge of the joint distribution of $(G_{N,0},G_{N,1},\ldots)$ would in principle characterize all coefficients in the corresponding expansions. 

\section{Macroscopic fluctuations of the stationary mass}
\label{sec:marginal_mass}

We now study the stationary mass at a prescribed observation site. Related proofs are discussed in  Appendix \ref{app:Wfield}. Since
\begin{equation}
\label{eq:log_mass_identity}
-\frac{1}{\beta}\log a_i(\omega)
=
V_i-m_N+\frac{1}{\beta}\log\!\left[Z_N(\beta)e^{\beta m_N}\right],
\end{equation}
the leading fluctuations at fixed $\beta>0$ are determined by the height of the potential above its global minimum.  
In fact, the last term in Eq.~\eqref{eq:log_mass_identity} is negligible on the diffusive scale, because
\begin{equation}
0\leq
\frac{1}{\beta\sqrt N}\log\!\left[Z_N(\beta)e^{\beta m_N}\right]
\leq
\frac{\log(N+1)}{\beta\sqrt N}
\xrightarrow[N\to\infty]{}0.
\end{equation}

Let $H_1$ and $H_2$ be independent standard half-Gaussian random variables,
\begin{equation}
\label{eq:half_gaussian_density}
\mathbb{P}(H_j\in dh)
=
\sqrt{\frac{2}{\pi}}e^{-h^2/2}\,\mathbf 1_{(0,\infty)}(h)\,dh,
\qquad j=1,2,
\end{equation}
where $\mathbf 1_{(a,b)}(h)$ is the characteristic function of the interval $(a,b)$.
For every fixed $x\in[0,1]$, the logarithmic mass at the macroscopic site $\lfloor xN\rfloor$ converges in distribution to the non-degenerate variable:
\begin{equation}
\label{eq:W_bulk_limit}
-\frac{\log a_{\lfloor xN\rfloor}(\omega)}{\beta\sqrt N}
\xrightarrow[N\to\infty]{d}
W(x)
:=
\max\!\left\{\sqrt{x}\,H_1,\sqrt{1-x}\,H_2\right\}.
\end{equation}

This variable describes the random coefficients of the stretched-exponential decay of the stationary mean masses.
The two terms in $W(x)$ arise from the minimum of the potential on the two sides of the observation point.  They are independent because they depend on disjoint sets of increments.  The law of $W(x)$ is symmetric under $x\leftrightarrow1-x$ and in particular interpolates continuously between a single half-Gaussian at the boundaries (i.e. $x\in\{0,1\}$) and the maximum of two equally weighted half-Gaussians at $x=1/2$. The boundary result extends in particular to any sequence of sites whose distance from one boundary is sublinear in $N$.  If $i_N/N\to0$, then
\begin{equation}
\label{eq:W_left_boundary_limit}
-\frac{\log a_{i_N}(\omega)}{\beta\sqrt N}
\xrightarrow[N\to\infty]{d} H_2,
\end{equation}
whereas, if $(N-i_N)/N\to0$, then
\begin{equation}
\label{eq:W_right_boundary_limit}
-\frac{\log a_{i_N}(\omega)}{\beta\sqrt N}
\xrightarrow[N\to\infty]{d} H_1.
\end{equation}
Thus the sublinear boundary regime is encoded in Eq.~\eqref{eq:W_bulk_limit} through $W(0)=H_2$ and $W(1)=H_1$.

\section{Localization at the deepest site}
\label{sec:localization}

Let $i_\star$ denote as in Eq. \eqref{eq:istar_definition_intro} the first site where the potential reaches its global minimum:
\begin{equation}
V_{i_\star} = \min_{0 \le j \le N} V_j .
\end{equation}
The extremal fractional mass is the stationary mass fraction residing on this deepest trap, denoted by $a_{i_\star} \equiv a_{i_\star}(\omega)$.

\subsection{High-disorder regime ($\beta>0$ and $N \to \infty$)}

For $\beta>0$ fixed, we can express this quantity as:
\begin{equation}
\label{eq:a_istari_def_main}
a_{i_\star} = \left(1 + \sum_{j=1}^{i_\star} e^{-\beta (V_{i_\star-j}-V_{i_\star})} + \sum_{j=1}^{N-i_\star} e^{-\beta (V_{i_\star+j}-V_{i_\star})}\right)^{-1}.
\end{equation}

Define the left and right barriers $L_j^\star=V_{i_\star-j}-V_{i_\star}$ and $R_j^\star=V_{i_\star+j}-V_{i_\star}$ on their respective available horizons. Since the first minimum is selected, $L_j^\star>0$ for $j\ge1$, whereas $R_j^\star\ge0$. Then
\begin{equation}\label{ExtremeFiniteMass}
a_{i_\star}=\frac{1}{1+\sum_{j=1}^{i_\star}e^{-\beta L_j^\star}+\sum_{j=1}^{N-i_\star}e^{-\beta R_j^\star}}.
\end{equation}
Appendix~\ref{app:minimum_decomposition} discusses the finite-horizon decomposition for random walk paths seen from its minimum and the related convergence to infinite-horizon paths conditioned to stay positive developed in\cite{bertoin1993,bertoin1994}. Such results offer a characterization of the infinite volume limit of the random variable $a_{i_\star}$ above. In particular, it is known that after diffusive scaling these infinite-horizon walks conditioned to stay positive converge to so-called three-dimensional Bessel processes~\cite{caravenna2008}. 

\subsection{Finite low-temperature limit ($\beta \to \infty$)}
At fixed $N$, the low-temperature expansion is read directly from the energy-level counts introduced in Eq.~\eqref{eq:energy_level_counts} and the related discussed expansion.  Indeed, since
\begin{equation}
 e^{\beta m_N}Z_N(\beta)=\sum_{r\ge0}G_{N,r}e^{-\beta r},
\end{equation}
for every selected minimizer $i_\star$ (as for $N$ finite this is not unique with positive probability) we have
\begin{equation}
\label{eq:minimum_mass_energy_levels}
a_{i_\star}
=\left(\sum_{r\ge0}G_{N,r}e^{-\beta r}\right)^{-1}
=\frac{1}{G_{N,0}}-\frac{G_{N,1}}{G_{N,0}^2}e^{-\beta}
+O(e^{-2\beta}).
\end{equation}
Thus the zero-temperature mass is shared equally among the sites attaining the global minimum, and $a_{i_\star}\to G_{N,0}^{-1}$ as $\beta\to\infty$.  
These variables $G_{N,r}$'s and their joint distribution might in principle be explicitly characterized for large $N$. For example, for the marginal $G_{N,0}$, 
it is possible to show that its limiting distribution is a geometric random variable on the positive integers with success parameter $1/2$, that is:
\begin{equation}
\label{eq:G0_geometric_main}
G_{N,0}\xrightarrow[N\to\infty]{d}
G_0,
\quad \text{with} \quad
\mathbb P(G_0=k)=2^{-k},\quad k\ge1,
\end{equation} 
see Appendix~\ref{app:energy_level_counts}. 

\subsection{Critical scaling regime ($\beta_N=\kappa/\sqrt{N}$)}
If the disorder strength is scaled diffusively as $\beta=\kappa/\sqrt{N}$ (for a fixed $\kappa>0$), the mass does not localize on a single site but spreads over a macroscopic portion of the system, decaying as $1/N$. More precisely , let $\beta=\kappa/\sqrt{N}$ with $\kappa>0$. As $N \to \infty$, the scaled variable $N a_{i_\star}$ converges in law to a functional of Brownian motion:
\begin{equation}
\label{eq:critical_maximum_limit}
N a_{i_\star} \xrightarrow[N \to \infty]{d} \Phi_\kappa \equiv \frac{1}{\int_0^1 e^{-\kappa(M - B_t)} dt}
\end{equation}
where $B_t$ is a standard Brownian motion on $[0,1]$ and $M = \max_{s \in [0,1]} B_s$.

\section{Arcsine laws and the disorder-averaged profile}
\label{sec:trapping}

The arcsine law enters the model in three complementary ways: through the position of the deepest valley, through the competition between the two sides of a macroscopic observation point, and through the disorder-averaged mass profile.  We state the three relations separately. Related proofs are presented in Appendix \ref{app:proofs_sec2}.

\subsection{Location of the deepest valley - (L\'evy's third arcsine law )}

Let $i_\star$ be the first site where $V$ reaches its global minimum on $[0,N]$, equivalently the first site where $S:=-V$ reaches its global maximum.  For centred independent increments with finite, non-zero variance, 
Donsker's invariance principle~\cite{donsker1951} and uniqueness of the time of the Brownian maximum on $[0,1]$ give
\begin{equation}
\label{eq:arcsine_location}
\frac{i_\star}{N}\xrightarrow[N\to\infty]{d}A_\star,
\end{equation}
where
\begin{equation}
\label{eq:arcsine_cdf_density}
\mathbb P(A_\star\le x):=\int_0^x f_{\rm arc}(z)\,dz=\frac{2}{\pi}\arcsin\sqrt{x},
\quad \text{with} \,
f_{\rm arc}(x):=\frac{\mathbf 1_{(0,1)}(x)}{\pi\sqrt{x(1-x)}},
\end{equation}
which is the classical L\'evy's arcsine law.

\subsection{Which side determines the macroscopic fluctuation?}

The two independent terms in $W(x)$ retain the location information of the limiting valley.  Indeed,
\begin{align}
\label{eq:arcsine_branch_probability}
\mathbb P\bigl(W(x)=\sqrt{x}H_1\bigr)
&=\mathbb P\left(\frac{H_2}{H_1}\le\sqrt{\frac{x}{1-x}}\right)\\
&=\frac{2}{\pi}\arctan\sqrt{\frac{x}{1-x}}
=\frac{2}{\pi}\arcsin\sqrt{x}.
\end{align}
Thus the probability that the left part of the landscape supplies the deeper contribution to $W(x)$ is exactly the probability that the Brownian localization centre lies to the left of $x$.

\subsection{Averaged stationary masses}

When averaging over the disorder, as anticipated in the asymptotics in Eq.~\eqref{AvArcsin} and discussed in the numerics in Fig.~\ref{fig:arcsin_profile}, the fraction of mass $\disavg{a_{\lfloor xN\rfloor}}$ at a given site at linear distance from the boundaries manifests again the arcsine signature. 
As discussed in Appendix~\ref{app:proofs_sec2}, this can already be seen by inspection of the Laplace transform of the macroscopic variables emerging in~\eqref{eq:intro_W_limit}, for which, for any fixed $x\in (0,1)$ as $t\to\infty$, one has the first order asymptotics

\begin{equation}
\label{LaplaceExpansion}
\mathbb E[e^{-tW(x)}]\sim \frac{4}{t^2} f_{\rm arc}(x).
\end{equation}

For the averaged masses appearing in  
 Eq.~\eqref{AvArcsin},  one can argue at fixed temperature $\beta>0$ by quenching the disorder $\omega$ and considering a random site $I_N \in \{0,1,\ldots, N\}$ sampled from the stationary Gibbs profile, that is:
\begin{equation}
\label{eq:gibbs_sampled_site}
\mathbb P_\omega(I_N=i):=a_i(\omega),\qquad 0\le i\le N.
\end{equation}
By the path decomposition at the minimum $i_\star$ of the potential, it is possible to show, see Appendix \ref{arcsine3},  that
\begin{equation}
\label{eq:gibbs_site_localizes_at_minimum}
\frac{I_N-i_\star}{N}\xrightarrow[N\to\infty]{d}0,
\end{equation}
where such a convergence in distribution is with respect to the probability including both the disorder and the conditional sampling in Eq.~\eqref{eq:gibbs_sampled_site}.  
The latter, together with the limiting arcsine law for $i_\star/N$ in Eq.~\eqref{eq:arcsine_location} gives
\begin{equation}
\label{eq:arcsine_limit_profile}
\frac{I_N}{N}\xrightarrow[N\to\infty]{d}A_\star,
\end{equation}

equivalently, for every bounded continuous test function $f$,
\begin{equation}
\label{eq:averaged_arcsine_weak}
\lim_{N\to\infty}\sum_{i=0}^N f(i/N)\disavg{a_i}
=\int_0^1\frac{f(x)}{\pi\sqrt{x(1-x)}}\,dx.
\end{equation}
This equivalent formulation of the convergence in \eqref{eq:arcsine_limit_profile} is nothing but the integral version of Eq.~\eqref{AvArcsin}, which would follow by arguing that one could differentiate the left and right hand sides above and get the same convergence for the local density, but we do not pursue this here.

\section{Numerics: mass profiles and inverse participation ratio}
\label{sec:numerics}

\begin{figure}[t]
\centering
\includegraphics[width=0.92\textwidth]{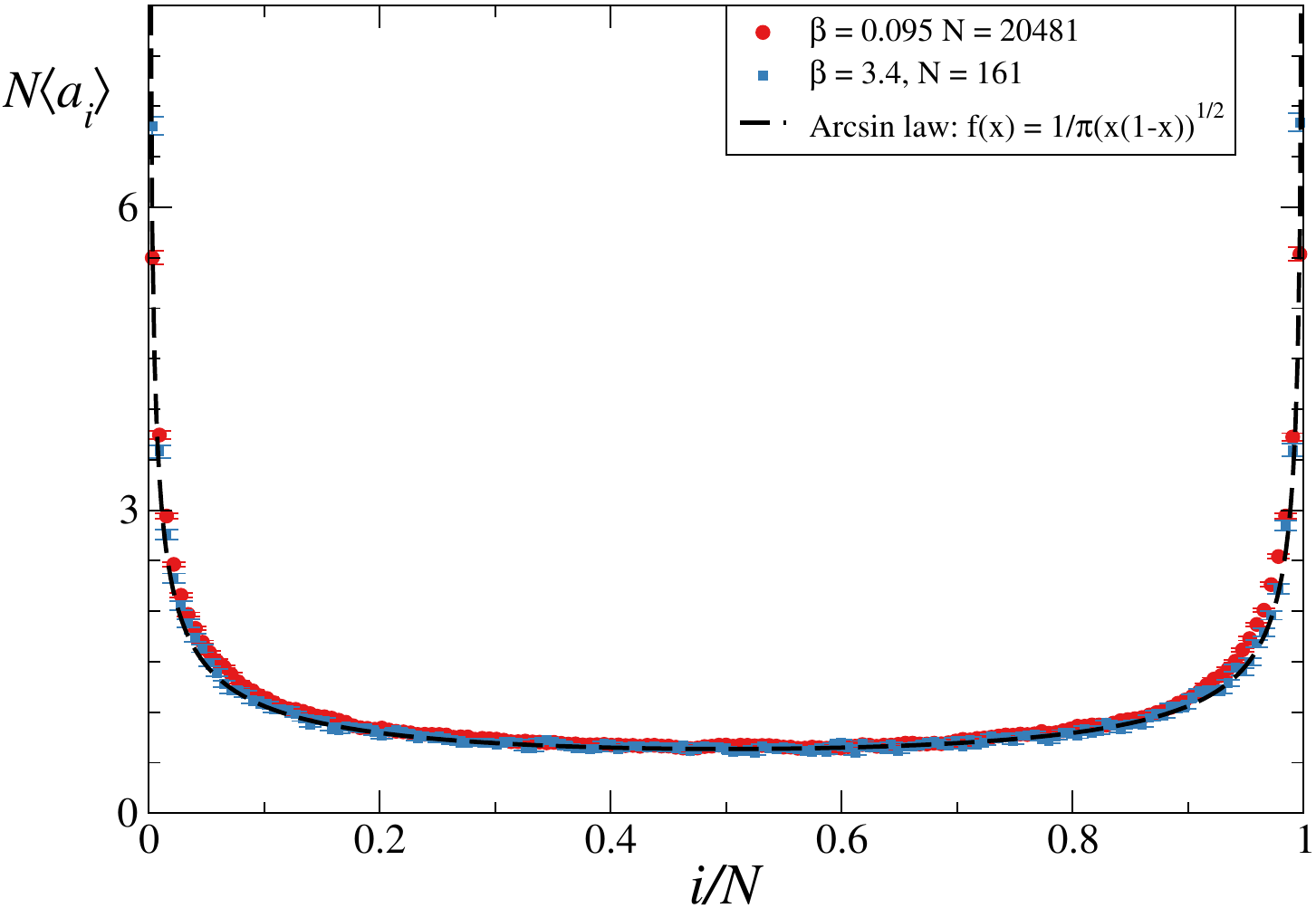}
\caption{Disorder-averaged stationary mass profile on the segment.  The plotted mass is normalised so that its sum over sites equals $N$, and therefore represents $N\disavg{a_i}$.  
Red circles correspond to $\beta=0.095$, $N=20481$, and blue squares to $\beta=3.4$, $N=161$.  The black dashed curve is the arcsine density $f_{\rm arc}(x)=[\pi\sqrt{x(1-x)}]^{-1}$.  Error bars show the numerical uncertainty of the disorder average.}
\label{fig:arcsin_profile}
\end{figure}

\begin{figure}[h]
\centering
\includegraphics[width=0.78\textwidth]{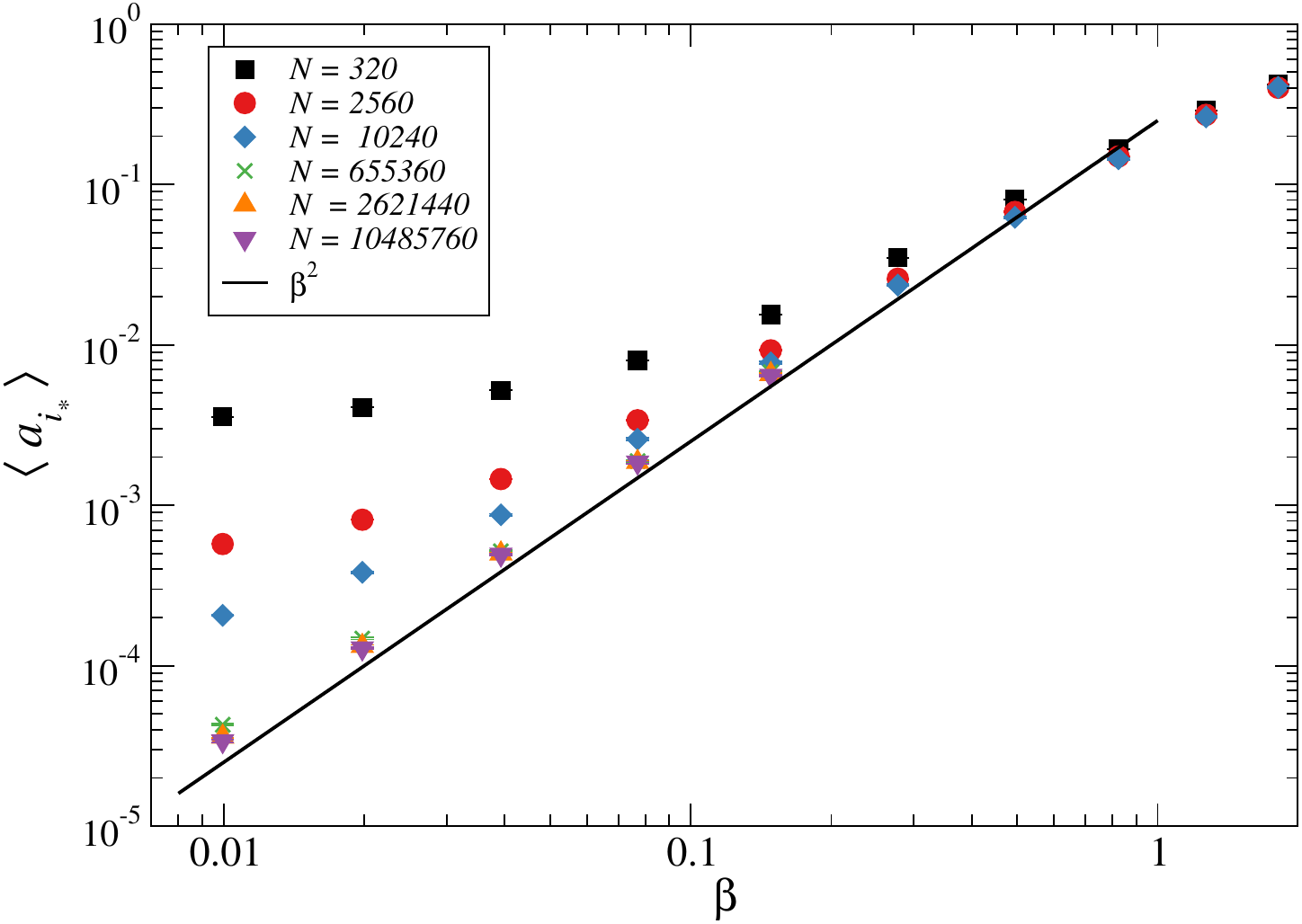}
	\caption{Mean mass at the localization peak as a function  of $\beta$ for samples of different sizes $N$.
	The straight line corresponds to the slope $\langle a_{{i_\star}}\rangle =\beta^2$
	(up to a multiplicative constant).}
\label{fig:mass_peak}
\end{figure}

\begin{figure}[t]
\centering
	\includegraphics[width=0.92\textwidth]{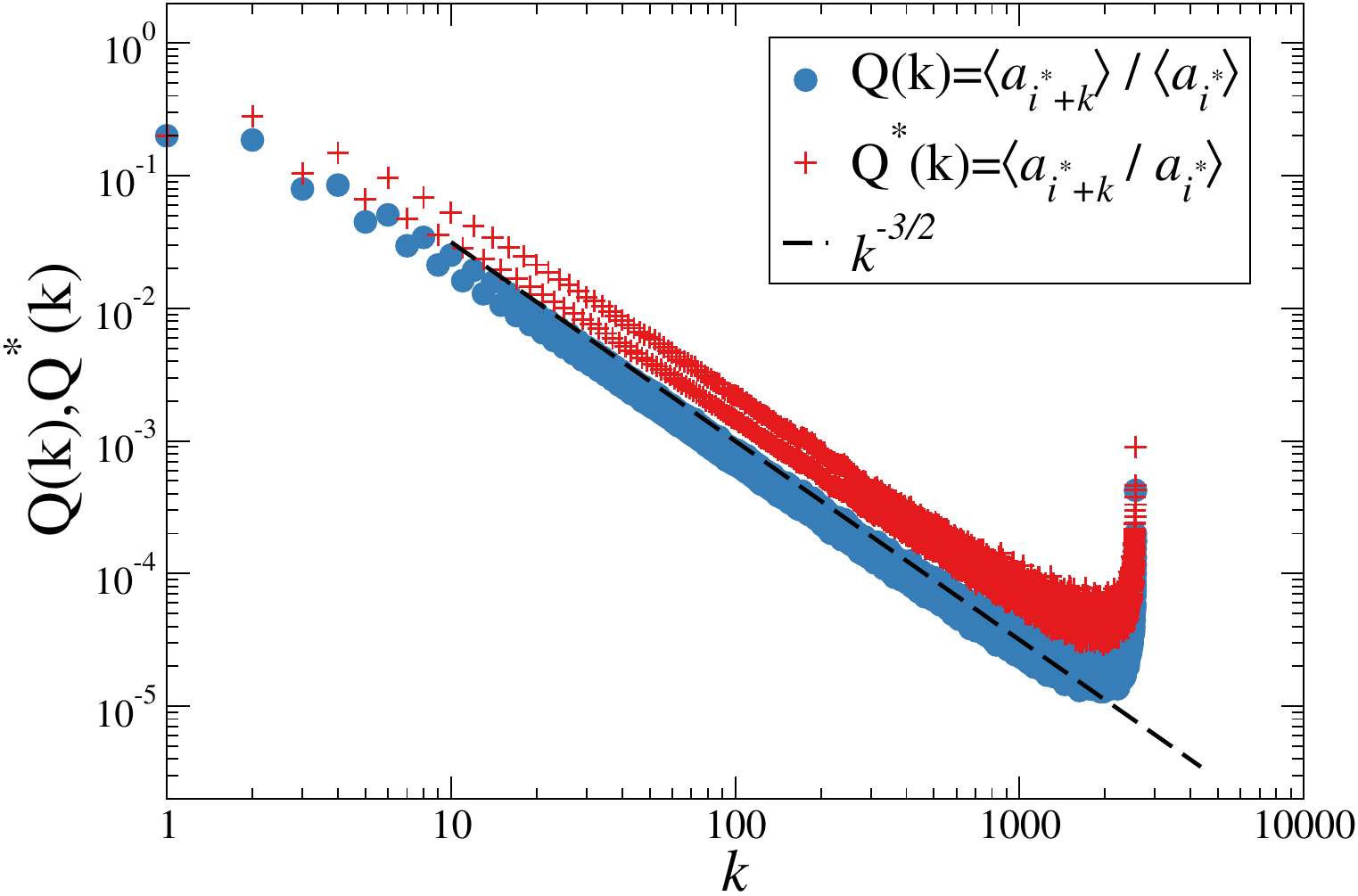}
	\caption{Plot of $Q(k)$ and $Q^*(k)$ for $\beta=\ln 5$ and N=2561.
	We also draw a $k^{-3/2}$ tail.}
	\label{fig:k3mezzi_profile}
\end{figure}

\begin{figure}[t]
\centering
	\includegraphics[width=0.92\textwidth]{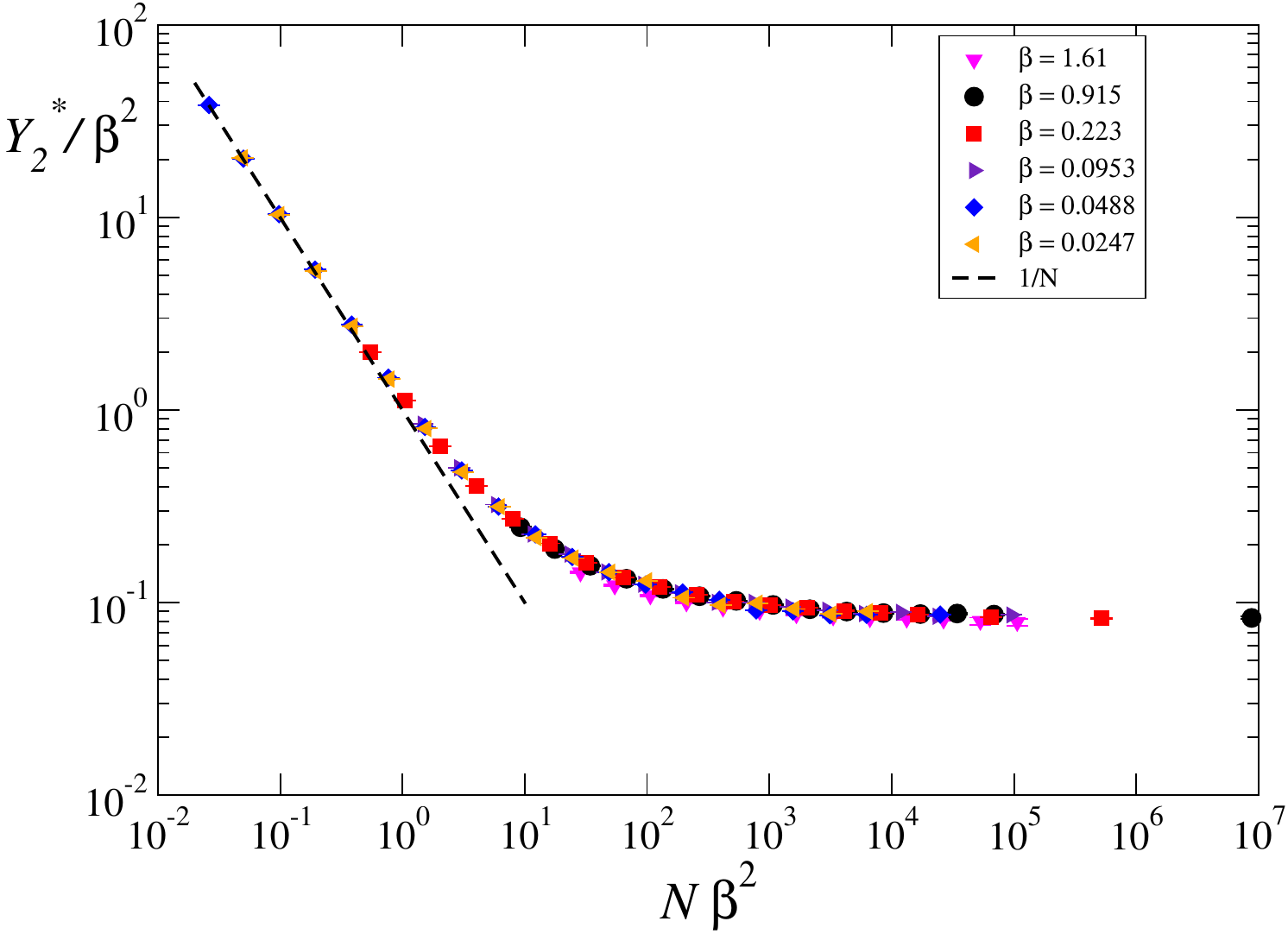}
	\caption{$Y_2^*(N,\beta)/\beta^2$ \textit{versus} $N\beta^2$.}
	\label{fig:Y2_KMP_beta}
\end{figure}

We now discuss some numerical results, useful to complement the analytical ones exposed in the previous sections.
We start with the already introduced Fig.~\ref{fig:arcsin_profile} for the arcsine profile and with
Fig.~\ref{fig:mass_peak}, plotting the mass fraction in the peak. 
Then, we are discussing the mass profile around the
peak, see Fig.~\ref{fig:k3mezzi_profile}, and the inverse participation ratio, the order parameter to study the
localization in proximity of the transition at $\beta=0$, see Fig.~\ref{fig:Y2_KMP_beta}.

\textit{The arcsine profile---}%
In Fig.~\ref{fig:arcsin_profile} we compare the arcsine density $f_{\rm arc}$ in (\ref{eq:arcsine_cdf_density}) 
with the numerical distribution of the 
normalized mass $N\langle a_i\rangle$. The agreement improves by increasing $N$, of course, 
but for fixed $N$ it also increases by increasing $\beta$, because the larger the $\beta$,
the greater the localization. 

\textit{The mass peak---}%
In Fig.~\ref{fig:mass_peak} we plot the average over disorder of the extremal mass as a function
of the control parameter $\beta$, for different values of $N$.
With increasing $N$ a square power law is identified, 
which is consistent with Eq.\eqref{ExtremeFiniteMass} and, as mentioned, the random walk paths $L^\star$ and $R^\star$ therein are known to be diffusive. Furthermore, for fixed $N$ 
when $\beta\to 0$ $\langle a_{{i_\star}}\rangle \to 1/(N+1)$.
In the opposite limit of large $\beta$, at fixed $N$, 
$\langle a_{{i_\star}}\rangle \to \langle G_{N,0}^{-1}\rangle$.  
Using Eq.~\eqref{eq:G0_geometric_main}, the subsequent large-$N$ limit is 
$\mathbb E[G_0^{-1}]=\log 2$. 

\textit{The mass profile---}%
Let us now discuss the mass profile around the peak.
The right-hand barrier at finite $N$ is $R_k^\star=V_{i_\star+k}-m_N$, defined for $k\le N-i_\star$. Appendix~\ref{app:minimum_decomposition} identifies its fixed-window limit as $X_k^\uparrow$. The continuum approximation below concerns this limiting conditioned walk. Appendix~\ref{app:mass_profile_peak} gives its exact lattice kernel and the corresponding large-$k$ density approximation
\begin{equation}
P(x,k)\simeq\sqrt{\frac2\pi}\frac{x^2}{k^{3/2}}e^{-x^2/(2k)},\qquad x>0,
\label{eq.PXx}.
\end{equation}

For finite samples, the following observables are understood conditionally on $i_\star+k\le N$. For fixed $k$ this restriction has probability tending to one as $N\to\infty$. In principle we are interested in the ratio
\be
Q(k) = \frac{\langle a_{{i_\star}+k}(\omega)\rangle}{\langle a_{{i_\star}}(\omega)\rangle},
\ee
but each average value would require the exact knowledge of the partition function.
Here we limit ourselves to evaluate the proxy
\be
Q^*(k) = \left\langle \frac{a_{{i_\star}+k}(\omega)}{a_{{i_\star}}(\omega)}\right\rangle,
\ee
and to compare numerically $Q(k)$ with $Q^*(k)$.

After taking $N\to\infty$ at fixed $k$, the proxy is $Q^*(k)=\mathbb E[e^{-\beta X_k^\uparrow}]$. 
The fixed-$\beta$ asymptotic, including lattice parity, is discussed in Appendix~\ref{app:mass_profile_peak}. 
Using Eq.~\eqref{eq.PXx} instead gives the continuum approximation below, appropriate to the weak-disorder crossover:

\bea
Q^*(k)=\mathbb E[e^{-\beta X_k^\uparrow}] &{\simeq}&
\int_0^\infty dx \sqrt{\frac{2}{\pi}} \frac{x^2}{k^{3/2}} e^{-x^2/2k} e^{-\beta x} \\
&=& \sqrt{\frac{2}{\pi}} \frac{1}{\beta^3 k^{3/2}} 
\int_0^\infty dy y^2 e^{-y} e^{-y^2/2\beta^2 k} \\
	&\sim & 
\begin{cases}
1 & \beta^2 k \ll 1 \\
\sqrt{\frac{8}{\pi}}	\frac{1}{\beta^3 k^{3/2}} & \beta^2 k \gg 1
\end{cases}
.
\eea

In order to highlight the algebraic decay in Fig.~\ref{fig:k3mezzi_profile} we
have used $\beta =\ln 5$. The $k^{-3/2}$ decay is visible, both for
$Q^*(k)$ and $Q(k)$.

Considering that $\langle a_{i_\star} \rangle \simeq c\beta^2$ for vanishing $\beta$, in this limit we expect that
\be
\langle a_{i_\star+k} \rangle \simeq
\begin{cases}
\beta^2 & k \ll 1/\beta^2 \\
        \frac{1}{\beta k^{3/2}} & k\gg 1/\beta^2 
\end{cases} .
\label{eq.aik}
\ee
Since $\int_{\beta^{-2}}^\infty dk/(\beta k^{3/2}) = O(1)$, the $k^{-3/2}$ tails 
carry a finite fraction of the mass.

\textit{The inverse participation ratio}---%
A localization transition is often characterized using the inverse participation ratio $Y_2$ as order parameter.
For a normalized total mass,
\be
Y_2(\beta,N) = \sum_{i=0}^N \langle \overline{c_i^2} \rangle ,
\ee
where the horizontal bar is the average over the invariant measure and $\langle\cdots\rangle$ means
the average over disorder. 

The quantities $a_i(\omega)=\bar c_i$ have been determined even without the knowledge of the invariant measure,
but the determination of $\overline{c_i^2}$ requires the explicit solution of a linear system of $N^2\times N^2$
equations because the evaluation of $\overline{c_i^2}$ implies the knowledge of $\overline{c_ic_j}$.
For small $N$ it is possible a numerical solution, but with increasing $N$ it quickly becomes unfeasible.

Therefore, rather than $Y_2$ we have studied
\be
Y_2^*(\beta,N) = \sum_{i=0}^N \langle (\overline{c_i})^2 \rangle =
\sum_{i=0}^N \langle a_i^2(\omega)\rangle,
\ee
which is a lower bound, $Y_2^*\le Y_2$, by Jensen's inequality.

In Fig.~\ref{fig:Y2_KMP_beta} we plot $Y_2^*(\beta,N)/\beta^2$ as a function of $N\beta^2$, 
showing that data sets for different $\beta$ collapse on a single curve.
This is because $Y_2^*(\beta,N)$ has different behaviors in the different regimes introduced 
in Sec.~\ref{sec:partition_function}: it decays as $1/N$ in the subcritical (homogeneous)
regime and it is of order $\beta^2$ in the supercritical (localized) regime.
The crossover between the two regimes occurs along the critical line $\beta^2 N$ constant.


\section{Discussion and outlook}
\label{sec:discussion}
We have studied a disordered, continuous mass-exchange model which is controlled 
by the geometry of a recurrent random landscape.
The mass-exchange dynamics is generically irreversible, although its stationary mean profile has a remarkably neat Gibbsian representation,  Eq.~\eqref{eq:central_gibbs_profile}.

\begin{figure}[t]
\centering
	\includegraphics[width=0.92\textwidth]{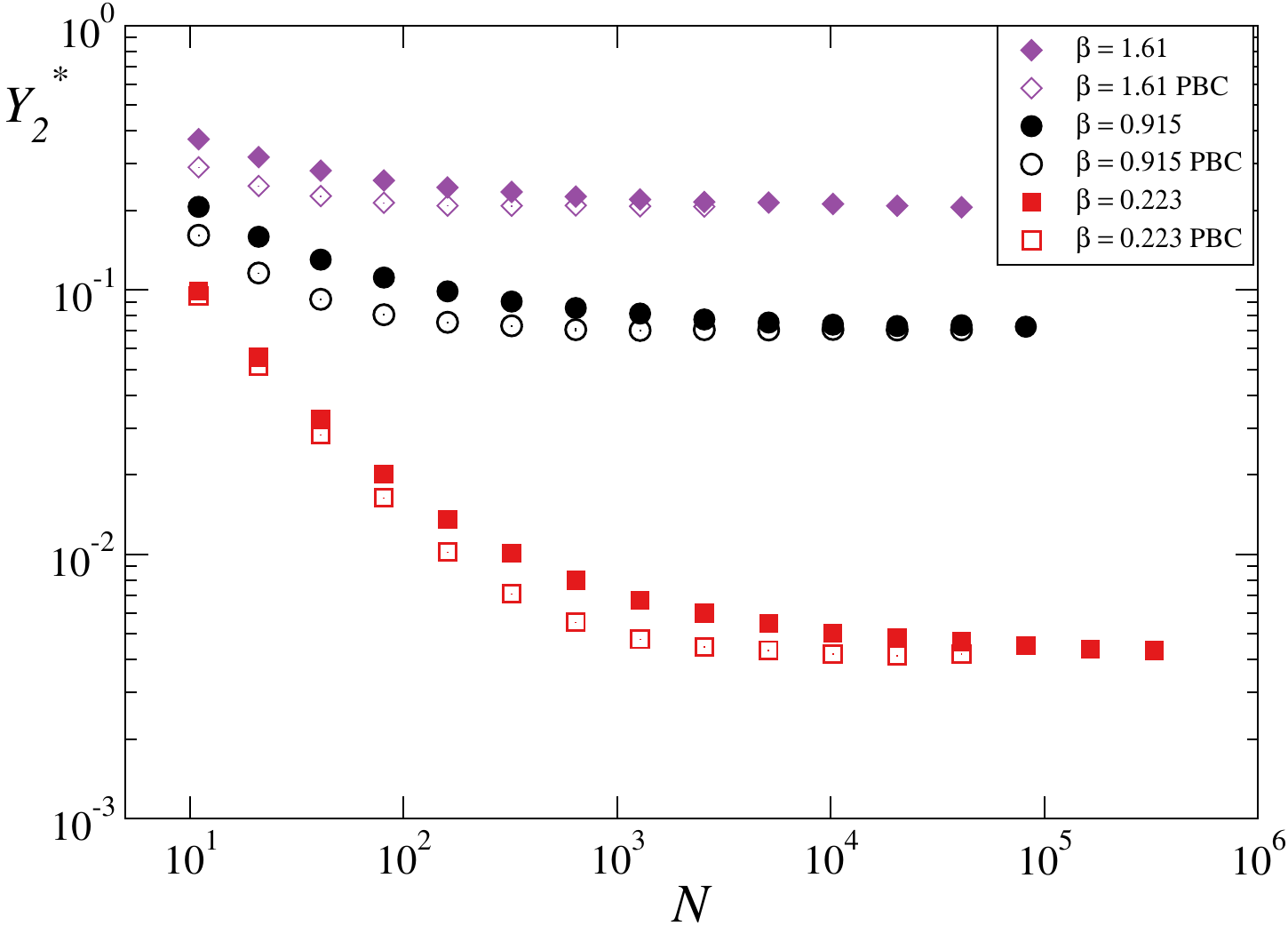}
	\caption{Comparison of $Y_2^*(N,\beta)$ for closed boundary conditions (full symbols)
	and periodic boundary conditions (open symbols).}
	\label{fig:Y2_PBC}
\end{figure}

We have discussed closed boundary conditions in detail and also studied the periodic model (Appendix~\ref{app:periodic}). On the ring a non-zero total potential increment generally produces a stationary circulating current. The profile then involves a current correction to the product formula. In the balanced case, when the product of the bond ratios is one, the current vanishes and a periodic Gibbs representation is recovered.
Figure~\ref{fig:Y2_PBC} compares $Y_2^*$ for closed and periodic boundaries. 
The numerical data suggest that for diverging $N$, $Y_2^*$ is insensitive to boundary conditions,
and that for periodic ones the asymptotic value is attained faster. 
The former evidence is raisonable, but the
equality of the limiting values is not established by the arguments given here.

The findings in here open up the way for a number of natural follow-up questions to be considered for the future.  
One may study quantitative convergence rates in the critical regime, 
dynamical aspects, characterization of the joint mass equilibrium distribution, and  
the extension of the analysis to non-Bernoulli bond disorder.  
Further directions include open boundary conditions, non-zero global drift, correlated environments 
and non-regular or other higher-dimensional-lattice geometries. 
Finally, we might also consider models where $c'_i$ is not a fraction of $(c_i + c_{i+1})$ but
a linear combination of $c_i$ and $c_{i+1}$~\cite{LPP}.

\begin{acknowledgements}
		PP acknowledges the financial support from the Italian MUR PRIN2022 project ‘Breakdown of ergodicity in classical and quantum many-body systems’ (BECQuMB) Grant No. 20222BHC9Z.
\end{acknowledgements}

\newpage
\appendix

\section{Partition function and local masses}
\label{app:partition_and_marginal}

\subsection{Stationary mean profile at fixed disorder}

The statements in Eqs.~\eqref{eq:stationary_mean_profile}--\eqref{eq:stationary_normalization}  follow by imposing  that at equilibrium the generator in Eq. \eqref{Generator} acting on the function giving the $i-$coordinate of the configuration $c$ must equal to zero. Let us spell this in physics terms.

Fix the environment, namely the collection of edge distributions $\{g_i\}_{i=0}^{N-1}$,
and let $m_i(t)=\mathbb{E}[c_i(t)]$, where the expectation is only over the mass-exchange
dynamics.  If edge $i$ rings with rate $\lambda_i>0$, the conditional mean of the
redistributed masses is
\begin{equation}
 \mathbb{E}[c'_i\mid c]=\overline{r_i}(c_i+c_{i+1}),
 \qquad
 \mathbb{E}[c'_{i+1}\mid c]=(1-\overline{r_i})(c_i+c_{i+1}).
\end{equation}
It is therefore convenient to introduce the mean current from site $i$ to site $i+1$,
\begin{equation}
 J_i(t)=\lambda_i\big[(1-\overline{r_i})m_i(t)
       -\overline{r_i}m_{i+1}(t)\big].
\end{equation}
The mean masses satisfy the continuity equations
\begin{equation}
 \frac{d m_0}{dt}=-J_0,\qquad
 \frac{d m_i}{dt}=J_{i-1}-J_i\quad (1\leq i\leq N-1),\qquad
 \frac{d m_N}{dt}=J_{N-1}.
\end{equation}
At stationarity the bulk equations imply that the current is constant along the chain,
whereas either boundary equation forces this constant to be zero.  Hence, writing
$a_i=\mathbb{E}_{\rm st}[c_i]$, for every $i=0,\ldots,N-1$ one has
\begin{equation}
 (1-\overline{r_i})a_i=\overline{r_i}a_{i+1},
 \qquad\text{or equivalently}\qquad
 a_{i+1}=\gamma_i a_i.
\end{equation}
This is the usual zero-current recursion for a birth-and-death process.  Iteration gives
\begin{equation}
 a_i=a_0\prod_{k=0}^{i-1}\gamma_k,
\end{equation}
and conservation of the total mass, normalized here as $\sum_{i=0}^N a_i=1$,
yields $a_0=Z^{-1}$ with $Z$ given by Eq.~\eqref{eq:stationary_normalization}.
This proves Eq.~\eqref{eq:stationary_mean_profile}.  Notice that the rates $\lambda_i$
drop out of the zero-current relation; in particular, the result holds for the uniform
edge-update rates used in the main text.

\subsection{Derivation of Eq.~\eqref{eq:partition_critical_limit}}\label{app:CriticalPartition}
Let $B_t^N=V_{\lfloor Nt\rfloor}/\sqrt N$ on $[0,1]$. By definition of Riemann integral, taking into account the endpoints of the related discrete sum, we can write
\begin{equation}
\frac1N Z_N\left(\frac\kappa{\sqrt N}\right)
=\int_0^1 e^{-\kappa B_t^N}\,dt
+\frac1N e^{-\kappa B_1^N}.
\label{StepApprox}
\end{equation}
By Donsker's theorem, $B^N \xrightarrow[N\to\infty]{d} B$ in $D[0,1]$ with the Skorokhod $J_1$ topology. The functional $b\mapsto\int_0^1e^{-\kappa b(t)}dt$ is continuous at continuous paths for this topology: convergence to such a path is uniform, and the exponential integrands then converge uniformly. The continuous mapping theorem applies because Brownian paths are almost surely continuous~\cite{billingsley2013}. Since $B_1^N$ is tight, the endpoint term in Eq.\eqref{StepApprox} is $O_{\mathbb P}(N^{-1})$ and tends to zero. This proves Eq.~\eqref{eq:partition_critical_limit}.

\subsection{Derivation of Eq.~\eqref{eq:partition_fixed_beta_limit} }\label{app:limitExtremeMass}
Let $\beta > 0$ be fixed and recall the definition of $i_\star$ in Eq.~\eqref{eq:istar_definition_intro}.
 The partition function can be decomposed by splitting the sum around the random global (first) minimum $i_\star$ :
\begin{equation}
Z_N(\beta) = e^{-\beta V_{i_\star}} \left( 1 + \sum_{j=1}^{i_\star} e^{-\beta(V_{i_\star-j} - V_{i_\star})} + \sum_{j=1}^{N-i_\star} e^{-\beta(V_{i_\star+j} - V_{i_\star})} \right)
\end{equation}
Recalling that $m_N = V_{i_\star}$, we multiply both sides by $e^{\beta m_N}$:
\begin{equation}
Z_N(\beta)e^{\beta m_N}=1+\sum_{j=1}^{i_\star}e^{-\beta L_j^\star}+\sum_{j=1}^{N-i_\star}e^{-\beta R_j^\star}
\end{equation}

The decomposition and convergence of these exponential series are discussed in Appendix~\ref{app:minimum_decomposition}. With its convention
\[
\mathcal Z_\infty^{(1)}(\beta)=\sum_{j\ge1}e^{-\beta L_j^\uparrow},\quad
\mathcal Z_\infty^{(2)}(\beta)=\sum_{j\ge0}e^{-\beta R_j^\uparrow},
\]
the origin is counted only once, and their sum is the limit in Eq.~\eqref{eq:partition_fixed_beta_limit}.

\subsection{Derivation of Eqs.~\eqref{eq:W_bulk_limit}--\eqref{eq:W_right_boundary_limit}}\label{app:Wfield}
Let $i\in\{0,\ldots,N\}$ and decompose the depth of the global minimum relative to $V_i$ into its left and right contributions,
\begin{equation}
\label{eq:left_right_depths}
L_{N,i}=V_i-\min_{0\leq j\leq i}V_j,
\qquad
R_{N,i}=V_i-\min_{i\leq j\leq N}V_j.
\end{equation}
Then
\begin{equation}
\label{eq:depth_max_decomposition}
V_i-m_N=\max\{L_{N,i},R_{N,i}\}.
\end{equation}
The variables $L_{N,i}$ and $R_{N,i}$ depend on disjoint families of increments and are therefore independent.  For $i=\lfloor xN\rfloor$, Donsker's invariance principle and the reflection principle give
\begin{equation}
\label{eq:left_right_half_gaussian_limit}
\left(\frac{L_{N,\lfloor xN\rfloor}}{\sqrt N},
      \frac{R_{N,\lfloor xN\rfloor}}{\sqrt N}\right)
\xrightarrow[N\to\infty]{d}
\left(\sqrt{x}\,H_1,\sqrt{1-x}\,H_2\right).
\end{equation}
Indeed, each component is distributed as the running maximum of a random walk of the corresponding length, and the maximum of a standard Brownian motion on $[0,1]$ has the standard half-Gaussian law.

From Eq.~\eqref{eq:log_mass_identity},
\begin{equation}
-\frac{\log a_i(\omega)}{\beta\sqrt N}
=
\frac{V_i-m_N}{\sqrt N}
+
\frac{1}{\beta\sqrt N}
\log\!\left[Z_N(\beta)e^{\beta m_N}\right].
\end{equation}
Since $1\leq Z_N(\beta)e^{\beta m_N}\leq N+1$, the second term converges uniformly to zero.  Combining Eqs.~\eqref{eq:depth_max_decomposition} and \eqref{eq:left_right_half_gaussian_limit} with the continuous mapping theorem proves Eq.~\eqref{eq:W_bulk_limit}.

If $i_N/N\to0$, then $L_{N,i_N}/\sqrt N\to0$ in probability, whereas $R_{N,i_N}/\sqrt N \xrightarrow[N\to\infty]{d} H_2$.  This proves Eq.~\eqref{eq:W_left_boundary_limit}.  The proof of Eq.~\eqref{eq:W_right_boundary_limit} is symmetric.

\section{Mass at the deepest site}
\label{app:proofs_sec1}

\subsection{The potential seen from its first minimum}
\label{app:minimum_decomposition}
We summarize in this section the construction of the two sides of the valley and their limiting characterization as derived in \cite{bertoin1993,bertoin1994}.Throughout, $i_\star$ is the first minimizer of $(V_j)_{0\le j\le N}$, $m_N=V_{i_\star}$, and $G_{N,0}$ is defined by Eq.~\eqref{eq:energy_level_counts}. We retain the notation from Section~\ref{sec:localization}:
\[
L_j^\star=V_{i_\star-j}-m_N\quad(0\le j\le i_\star),\qquad
R_j^\star=V_{i_\star+j}-m_N\quad(0\le j\le N-i_\star).
\]
Both paths start at zero. The first-minimum convention implies
\[
L_j^\star>0\quad(1\le j\le i_\star),\qquad
R_j^\star\ge0\quad(0\le j\le N-i_\star).
\]
In particular, no left-hand site other than the origin contributes to $G_{N,0}$. This strict/non-strict distinction will be kept at every step below.

\paragraph{Finite-horizon splitting.}
For the auxiliary simple symmetric walk $S$ under $\mathbb P_x$, started at $x$, put
\[
\mathcal C_h^+=\{S_1\ge0,\ldots,S_h\ge0\},\qquad
\mathcal C_h^>=\{S_1>0,\ldots,S_h>0\},
\]
with both events equal to the full space when $h=0$. Conditionally on $i_\star=k$, the law of $(L_j^\star)_{0\le j\le k}$ is $\mathbb P_0(\cdot\mid\mathcal C_k^>)$, and that of $(R_j^\star)_{0\le j\le N-k}$ is $\mathbb P_0(\cdot\mid\mathcal C_{N-k}^+)$. These two conditioned paths are independent. To see this, reverse the first $k$ increments of $V$ and change their signs. They again form independent symmetric signs, independent of the last $N-k$ increments. Moreover, $\{i_\star=k\}$ is exactly the intersection of strict positivity of the reversed first block and non-negativity of the second block. Each condition involves only its own block, so the conditional law factors. This is the discrete splitting at the infimum presented in Bertoin~\cite{bertoin1993}.

This statement is exact at finite $N$. Without conditioning on $i_\star$, the two pieces share random lengths and are not independent.

\paragraph{The infinite-horizon conditioned walk.}
Set $\tau_{-1}=\inf\{j\ge0:S_j=-1\}$ and $q_h(n)=\mathbb P_n(\tau_{-1}>h)$ for $n\ge0$. The finite-horizon conditioning analysis in Bertoin--Doney~\cite{bertoin1994} identifies the limit of $\mathbb P_0(\cdot\mid\mathcal C_h^+)$ on each fixed initial window as a Doob transform of the killed walk. In this simple symmetric case its harmonic function is
\[
h(n)=n+1,\qquad n\ge0.
\]
We denote this limiting walk by $X^\uparrow$, consistently with Appendix~\ref{app:mass_profile_peak}. Its transition probabilities are
\[
p^\uparrow(n,n+1)=\frac{n+2}{2(n+1)},\qquad
p^\uparrow(n,n-1)=\frac{n}{2(n+1)}\quad(n\ge1),\qquad p^\uparrow(0,1)=1.
\]
For completeness, the reflection principle gives $q_{h-M}(n)/q_h(0)\to n+1$ for fixed $M,n$. Thus the conditional probability of an admissible path $(s_0,\ldots,s_M)$, with $s_0=0$, is
\[
2^{-M}{\bf1}_{\{s_1,\ldots,s_M\ge0\}}
\frac{q_{h-M}(s_M)}{q_h(0)}
\longrightarrow
2^{-M}{\bf1}_{\{s_1,\ldots,s_M\ge0\}}h(s_M),
\]
which is exactly its Doob-transform probability.

The strictly positive conditioned walk has first step $1$. After that step, subtracting $1$ gives a walk starting at zero conditioned to stay non-negative for $h-1$ steps. Accordingly, take independent copies $X^\uparrow,\widetilde X^\uparrow$ and define the limiting half-paths by
\[
R_j^\uparrow=X_j^\uparrow\quad(j\ge0),\qquad
L_0^\uparrow=0,\quad L_j^\uparrow=1+\widetilde X_{j-1}^\uparrow\quad(j\ge1).
\]
These definitions account for both the shift and the strict positivity on the left.

The arcsine limit of $i_\star/N$ has no endpoint atoms, so both available horizons tend to infinity in probability. For each fixed $M$, the preceding conditional product laws therefore give
\[
\big((L_j^\star)_{j=0}^M,(R_j^\star)_{j=0}^M\big)
\xrightarrow{d}
\big((L_j^\uparrow)_{j=0}^M,(R_j^\uparrow)_{j=0}^M\big).
\]
The finite paths need only be defined on the event that their horizons exceed $M$, whose probability tends to one. Independence in the limit follows by mixing over $i_\star$: for $k,N-k\to\infty$ the two fixed-window conditional laws approach limits independent of $k$.
Caravenna--Chaumont~\cite{caravenna2008}, Theorem~1.1, gives the diffusive scaling limit of the infinite-horizon conditioned walk, namely a three-dimensional Bessel process in this finite-variance case. 

\paragraph{Control beyond a fixed window.}
Local path convergence must be supplemented by a uniform tail estimate to treat partition functions and total visit counts. By the Markov property, for $1\le j\le h$ and $n\ge0$,
\[
\mathbb P_0(S_j=n\mid\mathcal C_h^+)
=\frac{\mathbb P_0(S_j=n,\tau_{-1}>j)q_{h-j}(n)}{q_h(0)}.
\]
Reflection and binomial estimates give
\[
\mathbb P_0(S_j=n,\tau_{-1}>j)\le\frac{C(n+1)}{(j+1)^{3/2}},\quad
q_\ell(n)\le\frac{C(n+1)}{\sqrt{\ell+1}},\quad
q_h(0)\ge\frac{c}{\sqrt{h+1}}.
\]
These upper bounds remain valid when their right-hand sides exceed one. For fixed $\beta>0$, summing in $n$ gives
\[
\mathbb E_0[e^{-\beta S_j}\mid\mathcal C_h^+]
\le C_\beta\frac{\sqrt{h+1}}{(j+1)^{3/2}\sqrt{h-j+1}}.
\]
For $j\le h/2$, the square-root ratio is bounded. For $j>h/2$, summing the bound gives $C_\beta(h+1)^{-1/2}$. If $h>M$, these two estimates yield
\begin{equation}
\sup_{h\ge0}\mathbb E_0\!\left[\sum_{j=M+1}^h e^{-\beta S_j}\,\middle|\,\mathcal C_h^+\right]
\le\frac{C_\beta}{\sqrt{M+1}}.
\label{eq:uniform_valley_tail}
\end{equation}
For $h\le M$ the sum is empty. The left-hand walk satisfies the same estimate after its deterministic first step and the shift by one. For the limiting walk the Doob formula gives $\mathbb E[e^{-\beta X_j^\uparrow}]\le C_\beta(j+1)^{-3/2}$, so its exponential series is almost surely summable.

Define, with the origin counted only on the right,
\[
\mathcal Z_\infty^{(1)}(\beta)=\sum_{j\ge1}e^{-\beta L_j^\uparrow},\qquad
\mathcal Z_\infty^{(2)}(\beta)=\sum_{j\ge0}e^{-\beta R_j^\uparrow},\qquad
\mathcal Z_\infty(\beta)=\mathcal Z_\infty^{(1)}(\beta)+\mathcal Z_\infty^{(2)}(\beta).
\]
For a fixed truncation $M$, convergence follows from the joint local limit. Equation~\eqref{eq:uniform_valley_tail} then removes the truncation, by Markov's inequality, and proves
\[
e^{\beta m_N}Z_N(\beta)\xrightarrow{d}\mathcal Z_\infty(\beta),\qquad
a_{i_\star}\xrightarrow{d}\mathcal Z_\infty(\beta)^{-1}.
\]
This proves Eq.~\eqref{eq:partition_fixed_beta_limit} and identifies the limiting mass in Eq.~\eqref{eq:a_istari_def_main}.

\paragraph{The lowest energy-level count.}
\label{app:energy_level_counts}
At finite $N$, strict positivity on the left gives the exact identity
\[
G_{N,0}=\#\{0\le j\le N:V_j=m_N\}
=\sum_{j=0}^{N-i_\star}{\bf1}_{\{R_j^\star=0\}}.
\]
For each fixed $M$, its truncation converges to $\sum_{j=0}^M{\bf1}_{\{X_j^\uparrow=0\}}$. Here the finite-window paths take values in a discrete set, so the visit-count functional is continuous on their support. Since ${\bf1}_{\{S_j=0\}}\le e^{-\beta S_j}$ on a non-negative path, Eq.~\eqref{eq:uniform_valley_tail} also controls the omitted visits, uniformly in the finite horizon. It follows that
\[
G_{N,0}\xrightarrow{d}G_0:=\sum_{j\ge0}{\bf1}_{\{X_j^\uparrow=0\}}.
\]
To compute its law, start $X^\uparrow$ at zero. Its first step is deterministically $1$. For the Doob transform started at $1$, telescoping the harmonic factors along paths stopped at their first return to zero gives
\[
\mathbb P_1^\uparrow(\tau_0<\infty)
=\frac{h(0)}{h(1)}\mathbb P_1(\tau_0<\infty)=\frac12,
\]
where the final equality uses recurrence of the ordinary simple symmetric walk. After each return to zero, the strong Markov property restarts the same experiment. Thus
\[
\mathbb P(G_0\ge r)=2^{-(r-1)},\qquad
\mathbb P(G_0=r)=2^{-r},\qquad r\ge1.
\]
Consequently
\begin{equation}
G_{N,0}\xrightarrow{d}G_0,\qquad
G_0\sim\mathrm{Geom}(1/2),\qquad \mathbb E[G_0]=2,
\label{eq:G0_geometric_appendix}
\end{equation}
with support $\{1,2,\ldots\}$. This proves Eq.~\eqref{eq:G0_geometric_main}. Higher energy-level counts may be studied using the same two half-paths, but their joint law is not needed here.

\subsection{Derivation of Eq.~\eqref{eq:critical_maximum_limit}}
Let $S_i=-V_i$, $\beta_N=\kappa/\sqrt N$, and define $b_N(t)=S_{\lfloor Nt\rfloor}/\sqrt N$ for $0\le t\le1$. Donsker's theorem gives $b_N\xrightarrow[N\to\infty]{d} B$ in the cadlag space $D[0,1]$ with the Skorokhod $J_1$ topology. Since $i_\star$ is a maximizer of $S$, putting $M_N=\max_{0\le t\le1}b_N(t)$ gives
\[
N a_{i_\star}
=\left[\frac1N\sum_{i=0}^N
 e^{-\kappa(M_N-S_i/\sqrt N)}\right]^{-1}.
\]
Here the negative sign follows from $a_{i_\star}^{-1}=\sum_i e^{-\beta_N(V_i-m_N)}$, and $V_i-m_N=S_{i_\star}-S_i$.

For a bounded c\`adl\`ag path $b$, define explicitly
\[
H_\kappa(b)=\int_0^1\exp\{-\kappa[\sup_{0\le s\le1}b(s)-b(t)]\}\,dt.
\]
This functional is continuous at every continuous path for the $J_1$ topology. Indeed, $J_1$ convergence $b_n\to b$ to a continuous $b$ implies $\|b_n-b\|_\infty\to0$. Also $|\sup b_n-\sup b|\le\|b_n-b\|_\infty$. The exponents therefore converge uniformly and stay bounded, proving $H_\kappa(b_n)\to H_\kappa(b)$. Since Brownian paths admits an almost surely continuous version, the continuous mapping theorem applies~\cite{billingsley2013}. 

The sum and the integral differ by exactly the endpoint contribution:
\[
\frac1N\sum_{i=0}^N e^{-\kappa(M_N-S_i/\sqrt N)}
=H_\kappa(b_N)+\frac1N e^{-\kappa(M_N-b_N(1))}.
\]
The last term lies between zero and $1/N$, so it tends to zero deterministically. Therefore the denominator converges to $H_\kappa(B)$. This is finite and strictly positive, so inversion is continuous at its value and yields
\[
N a_{i_\star}\xrightarrow{d}
\Phi_\kappa=\left[\int_0^1e^{-\kappa(M-B_t)}\,dt\right]^{-1},
\qquad M=\sup_{0\le t\le1}B_t.
\]

\section{Arcsine laws and averaged profiles}
\label{app:proofs_sec2}

\subsection{Derivation of the arcsine law for $i_\star/N$}
The finite splitting of Appendix~\ref{app:minimum_decomposition} gives exactly
\[
\mathbb P(i_\star=k)=p_k^>p_{N-k}^+,\qquad 0\le k\le N,
\]
where $p_h^+=\mathbb P_0(\mathcal C_h^+)$, $p_h^>=\mathbb P_0(\mathcal C_h^>)$, with both probabilities equal to one at $h=0$. The strict condition on the left is essential because we select the first minimum. For the simple symmetric walk, reflection gives
\[
p_h^+=2^{-h}\binom{h}{\lfloor h/2\rfloor},\qquad
p_h^>=\frac12 p_{h-1}^+\quad(h\ge1).
\]
The second identity follows by forcing the first step to $1$ and shifting the remaining path by $1$. Stirling's formula then gives
\[
p_h^+\sim\sqrt{\frac{2}{\pi h}},\qquad
p_h^>\sim\frac1{\sqrt{2\pi h}}.
\]
Consequently, uniformly for $k/N$ in a compact subinterval of $(0,1)$,
\[
N\mathbb P(i_\star=k)\longrightarrow
\frac1{\pi\sqrt{x(1-x)}}\quad\hbox{when } k/N\to x.
\]
To extend this bulk approximation to weak convergence on $[0,1]$, the same binomial estimates give
\[
\mathbb P(i_\star=k)\le\frac{C}{\sqrt{(k+1)(N-k+1)}}.
\]
Summing near either endpoint shows that their probabilities are $O(\sqrt\delta)+o(1)$ for windows of length $\delta N$. A Riemann sum on $[\delta,1-\delta]$, followed by $\delta\downarrow0$, proves Eq.~\eqref{eq:arcsine_location} with density
\[
f_{\rm arc}(x)=\frac{{\bf1}_{(0,1)}(x)}{\pi\sqrt{x(1-x)}}.
\]
This also proves the horizon divergence used in Appendix~\ref{app:minimum_decomposition} directly, without using conditioned-walk limits.

\subsection{Derivation of Eq.~\eqref{eq:arcsine_branch_probability}}

Let $Z_1,Z_2$ be independent standard Gaussian variables, so that $H_j=|Z_j|$.  The ratio $Z_2/Z_1$ is a standard Cauchy variable and, after taking absolute values,
\begin{equation}
\mathbb P\left(\frac{H_2}{H_1}\le u\right)
=\frac{2}{\pi}\arctan u,
\qquad u\ge0.
\end{equation}
Taking $u=\sqrt{x/(1-x)}$ and using
$\arctan\sqrt{x/(1-x)}=\arcsin\sqrt{x}$ proves Eq.~\eqref{eq:arcsine_branch_probability}.

\subsection{Averaged limiting masses in Eq.~\eqref{eq:arcsine_limit_profile} and~\eqref{LaplaceExpansion}}\label{arcsine3}

Let us start with the asymptotics in Eq.~\eqref{LaplaceExpansion}.
For $w\ge0$, independence of the two half-Gaussian variables gives
\begin{equation}
\label{eq:W_distribution_appendix}
F_{W(x)}(w):=\mathbb P(W(x)\le w)=\mathbb P(\sqrt{x}H_1\le w) \mathbb P(\sqrt{1-x}H_2\le w) 
\end{equation}
$$=\left[2\Phi\left(\frac{w}{\sqrt{x}} \right)-1\right]\left[2\Phi\left(\frac{w}{\sqrt{1-x}} \right)-1\right]
=
\erf\left(\frac{w}{\sqrt{2x}}\right)
 \erf\left(\frac{w}{\sqrt{2(1-x)}}\right).
$$
The Taylor expansion
\begin{equation}
\erf z=\frac{2}{\sqrt\pi}\left(z-\frac{z^3}{3}+\mathcal O(z^5)\right)
\end{equation}
therefore yields
\begin{equation}
\label{eq:W_quadratic_cdf}
F_{W(x)}(w)=\frac{2w^2}{\pi\sqrt{x(1-x)}}+\mathcal O(w^4),
\qquad w\downarrow0,
\end{equation}
and by differentiating we obtain the density expansion
\begin{equation}
\label{eq:W_linear_density}
f_{W(x)}(w)=\frac{4w}{\pi\sqrt{x(1-x)}}+\mathcal O(w^3).
\end{equation}
It then follows by Laplace's method that, as $t\to\infty$,
\begin{align}
\mathbb E[e^{-tW(x)}]
&=\int_0^\infty e^{-tw}f_{W(x)}(w)\,dw\\
&=\frac{4}{\pi\sqrt{x(1-x)}}\frac1{t^2}
+\mathcal O(t^{-4}).
\label{eq:W_laplace_asymptotic}
\end{align}

We now show Eq.~\eqref{eq:arcsine_limit_profile}. Fix $\beta>0$.  Conditionally on the environment $\omega$, consider the variable $I_N$ as defined in \eqref{eq:gibbs_sampled_site} and notice that 
for every integer $j$ with $-i_\star\le j\le N-i_\star$, we can write
 the Gibbs profile relative to the minimum as
\begin{equation}
\label{eq:gibbs_profile_from_minimum_appendix}
a_{i_\star+j}
=\frac{e^{-\beta(V_{i_\star+j}-V_{i_\star})}}
{\displaystyle\sum_{\ell=-i_\star}^{N-i_\star}
 e^{-\beta(V_{i_\star+\ell}-V_{i_\star})}}.
\end{equation}  

By now using the limiting half-paths $L^\uparrow,R^\uparrow$ defined in Appendix~\ref{app:minimum_decomposition}, we can write
\begin{equation}
\mathcal Z_\infty(\beta)=1+\sum_{j\ge1}e^{-\beta L_j^\uparrow}+\sum_{j\ge1}e^{-\beta R_j^\uparrow},
\end{equation}
and the bound after Eq.\eqref{eq:uniform_valley_tail} guarantees summability end hence
\begin{equation}
\label{eq:infinite_valley_tail}
\frac{\sum_{j>M}e^{-\beta L_j^\uparrow}+\sum_{j>M}e^{-\beta R_j^\uparrow}}{\mathcal Z_\infty(\beta)}\longrightarrow0\quad\hbox{almost surely}.
\end{equation}
More importantly, its finite-horizon version and the fact that the Gibbs denominator is at least one give directly
\[
\mathbb P(|I_N-i_\star|>M)\le\frac{C_\beta}{\sqrt{M+1}}.
\]
Thus, for every $\eta>0$, some $M<\infty$ satisfies
\begin{equation}
\label{eq:gibbs_displacement_tightness}
\limsup_{N\to\infty}\mathbb P(|I_N-i_\star|>M)<\eta.
\end{equation}

Now fix $\varepsilon>0$.  For $N>M/\varepsilon$,
\[
\{|I_N-i_\star|>\varepsilon N\}
\subseteq
\{|I_N-i_\star|>M\}.
\]
Taking first $N\to\infty$ and then $M\to\infty$ proves Eq.~\eqref{eq:gibbs_site_localizes_at_minimum}.

Finally,
\[
\frac{I_N}{N}
=\frac{i_\star}{N}+\frac{I_N-i_\star}{N}.
\]
The first term converges in distribution to the arcsine variable $A_\star$, while the second converges to zero in probability.  Slutsky's theorem therefore yields the 
\[
\frac{I_N}{N}\xrightarrow[N\to\infty]{d}A_\star,
\]
which proves Eq.~\eqref{eq:arcsine_limit_profile}.

\section{Mass profile around the peak}
\label{app:mass_profile_peak}

We discuss here the $k^{-3/2}$ decay  in Fig.\ref{fig:k3mezzi_profile} using the conditioned-walk construction of Appendix~\ref{app:minimum_decomposition}. The finite right-hand potential is $R_k^\star=V_{i_\star+k}-m_N$, for $0\le k\le N-i_\star$. Its infinite-horizon limit is $R_k^\uparrow=X_k^\uparrow$, where $X^\uparrow$ is the Doob transform with harmonic function $h(n)=n+1$ defined there. In this appendix we first take the infinite-horizon limit and then study $k\to\infty$. The finite-horizon splitting and the geometric law for $G_{N,0}$ are proved entirely in Appendix~\ref{app:minimum_decomposition}.

For the simple symmetric random walk killed when it enters $(-\infty,-1]$, the
positive harmonic function is
\[
h(n)=n+1,\qquad n\geq0.
\]
Thus $X^\uparrow$ is the Doob $h$-transform of the killed walk.  Starting from
$X^\uparrow(0)=0$, for $n\geq0$ and $n\equiv k\pmod 2$,
\begin{equation}
\mathbb P\bigl(X_k^\uparrow=n\bigr)
=(n+1)\,\mathbb P_0\bigl(S_k=n,\tau_{-1}>k\bigr),
\label{eq:doob_mass_profile}
\end{equation}
where $\tau_{-1}=\inf\{j\geq0:S_j=-1\}$.  By the reflection principle,
\[
\mathbb P_0\bigl(S_k=n,\tau_{-1}>k\bigr)
=\mathbb P_0(S_k=n)-\mathbb P_0(S_k=n+2).
\]
Consequently,
\begin{equation}
\mathbb P\bigl(X_k^\uparrow=n\bigr)
=\frac{2(n+1)^2}{k+n+2}\,
2^{-k}\binom{k}{(k+n)/2},
\qquad n\equiv k\pmod2.
\label{eq:conditioned_kernel_exact}
\end{equation}
This identity is exact; in particular, no continuum approximation is involved at
this stage.

The local central limit theorem applied to Eq.~\eqref{eq:conditioned_kernel_exact}
gives, 
when $n/\sqrt{k}$ ranges in a compact subset of $(0,\infty)$,

\begin{equation}
\mathbb P\bigl(X_k^\uparrow=n\bigr)
\sim \sqrt{\frac{8}{\pi}}\,
\frac{n^2}{k^{3/2}}\exp\left\{-\frac{n^2}{2k}\right\},
\label{eq:conditioned_kernel_lclt}
\end{equation}
on the parity class $n\equiv k\pmod2$.  Equivalently, replacing the lattice sum by
a continuum integral with lattice spacing two gives the density
\begin{equation}
P(x,k)\simeq \sqrt{\frac{2}{\pi}}\,
\frac{x^2}{k^{3/2}}\exp\left\{-\frac{x^2}{2k}\right\}.
\label{eq.PXx_appendix}
\end{equation}


\section{Periodic boundary conditions}
\label{app:periodic}
On the ring we use the same sites $0,\ldots,N$, now with an additional bond $(N,0)$. There are $N+1$ bonds, indexed $i=0,\ldots,N$, with $i+1$ understood modulo $N+1$. Bond updates have unit rates. The splitting means and ratios remain $\overline{r_i}$ and $\gamma_i=(1-\overline{r_i})/\overline{r_i}$. Define $J$ as the stationary current from $i$ to $i+1$. It is independent of $i$ by stationarity and satisfies
\begin{equation}
J=(1-\overline{r_i})a_i-\overline{r_i}a_{i+1},
\qquad a_{N+1}=a_0.
\label{eq:pbc1}
\end{equation}
Thus
\begin{equation}
a_{i+1}=\gamma_i a_i-J(\gamma_i+1).
\label{eq:pbc2}
\end{equation}
For $0\le i\le N+1$ put
\[
P_0=1,\quad D_0=0,\qquad
P_i=\prod_{k=0}^{i-1}\gamma_k,\quad
D_i=\sum_{j=0}^{i-1}(\gamma_j+1)\prod_{q=j+1}^{i-1}\gamma_q\quad(i\ge1).
\]
Empty products are one. Iterating Eq.~\eqref{eq:pbc2} yields
\begin{equation}
a_i=P_i a_0-JD_i.
\label{eq:pbc3}
\end{equation}
Periodicity implies $J=\widetilde J a_0$, where
\[
\widetilde J=\frac{P_{N+1}-1}{D_{N+1}}.
\]
Mass conservation gives the explicit solution
\[
a_i=\frac{P_i-\widetilde J D_i}
{\sum_{\ell=0}^N(P_\ell-\widetilde J D_\ell)},\qquad 0\le i\le N.
\]
All numerators are positive. Indeed, with $T_i=D_i/P_i$ one has $0\le T_i\le T_{N+1}$ and $\widetilde J=(1-P_{N+1}^{-1})/T_{N+1}$; hence $1-\widetilde J T_i>0$ whether $P_{N+1}$ is above or below one.

For the sign disorder, $\gamma_i=\gamma^{\omega_i}$ with $\omega_i\in\{\pm1\}$ on all $N+1$ bonds. Keep $V_0=0$ and $V_i=-\sum_{k=0}^{i-1}\omega_k$ for $1\le i\le N+1$. Then $P_i=\gamma^{-V_i}$ and
\[
T_i=\sum_{j=0}^{i-1}\gamma^{V_{j+1}}(1+\gamma^{\omega_j}),\qquad T_0=0.
\]
The same profile can therefore be written as
\[
a_i=\frac{\gamma^{-V_i}\left[1-(1-\gamma^{V_{N+1}})T_i/T_{N+1}\right]}
{\sum_{\ell=0}^N\gamma^{-V_\ell}\left[1-(1-\gamma^{V_{N+1}})T_\ell/T_{N+1}\right]}.
\]
The current is zero exactly when $P_{N+1}=1$, equivalently $\prod_{i=0}^N\gamma_i=1$. For $\gamma>1$ this is $\sum_{i=0}^N\omega_i=0$, or $V_{N+1}=0$. In that case the profile reduces to the Gibbs formula. For $\gamma=1$ it is flat and the current vanishes for every sign realization.

\newpage


\end{document}